\documentclass[letterpaper,11pt]{article}
\usepackage[margin=1in]{geometry}

\usepackage[utf8]{inputenc} 
\usepackage[T1]{fontenc}    
\usepackage{hyperref}       
\usepackage{url}            
\usepackage{booktabs}       
\usepackage{amsfonts}       
\usepackage{nicefrac}       
\usepackage{microtype}      
\usepackage{graphicx}
\usepackage{amssymb,amsmath,amsthm,amsfonts}
\usepackage{dsfont}
\usepackage{float}
\usepackage{hyperref}
\usepackage{url}
\usepackage{stmaryrd}

\usepackage[noend]{algpseudocode}
\usepackage{algorithm}

\makeatletter
\renewcommand*\env@matrix[1][\arraystretch]{%
  \edef\arraystretch{#1}%
  \hskip -\arraycolsep
  \let\@ifnextchar\new@ifnextchar
  \array{*\c@MaxMatrixCols c}}
\makeatother

\algdef{SE}[SUBALG]{Indent}{EndIndent}{}{\algorithmicend\ }%
\algtext*{Indent}
\algtext*{EndIndent}

\DeclareMathOperator{\atanh}{atanh}
\DeclareMathAlphabet{\varmathbb}{U}{bbold}{m}{n}

\renewcommand{\P}{{\rm P}}

\newcommand{\B}{{\rm B}}
\renewcommand{\Pr}{\mathbb{P}}
\newcommand{\Q}{{\rm Q}}
\newcommand{\U}{{\rm U}}
\newcommand{\E}{\mathbb{E}}

\renewcommand{\H}{{\rm H}}
\newcommand{\I}{{\rm I}}
\newcommand{\Tr}{{\rm Tr}}

\newcommand{\trans}[1]{{#1}^\intercal}
\def\ind{{\mathbf{1}}}

\newtheorem{theorem}{Theorem}[section]
\newtheorem{lemma}[theorem]{Lemma}
\newtheorem{proposition}[theorem]{Proposition}
\newtheorem{corollary}[theorem]{Corollary}

\makeatletter
\let\@fnsymbol\@arabic
\makeatother

\begin{document}


\title{Spectral Bounds for the Ising Ferromagnet \\ on an Arbitrary Given Graph 
}




\author{Alaa Saade\thanks{
Laboratoire de Physique Statistique (CNRS UMR-8550),
  PSL Universit\'es \& \'Ecole Normale Sup\'erieure, 75005
  Paris 
}\and Florent Krzakala$^{1,}$\footnote{
Sorbonne Universit\'es, UPMC Univ. Paris 06
}$^{\ ,}$\footnote{Simons Institute for the Theory of Computing, University of California, Berkeley, Berkeley, CA,  94720
}\and Lenka Zdeborov\'a$^{3,}$\thanks{Institut de Physique Th\'eorique, CNRS, CEA, Universit\'e Paris-Saclay, 91191, Gif-sur-Yvette, France.
}
}
\date{}

\maketitle
\begin{abstract}
We revisit classical bounds of M. E. Fisher on the ferromagnetic Ising model \cite{fisher1967critical}, and show how to efficiently use them on an arbitrary given graph to rigorously upper-bound
  the partition function, magnetizations, and correlations. The results are valid on
  any finite graph, with arbitrary topology and arbitrary positive couplings and fields. Our results are
  based on high temperature expansions of the aforementioned quantities, and are expressed in terms of two related linear
  operators: the non-backtracking operator and the Bethe Hessian. As a by-product, we show that in a well-defined high-temperature region,
  the susceptibility propagation algorithm \cite{mezard2009constraint} converges and provides an upper bound on
  the true spin-spin correlations.
\end{abstract}

\section{Introduction}
\label{intro}
Undirected graphical models, or Markov random fields,  are a powerful paradigm for multivariate statistical modeling. They allow to encode information about the conditional dependencies of a large number of interacting variables in a compact way, and provide a unified view of inference and learning problems in areas as diverse as statistical physics, computer vision, coding theory or machine learning (see e.g. \cite{WJ2008,koller2009probabilistic} for examples of applications). 
Arguably one of the oldest and most prototypical example of a (pairwise) Markov random field is the Ising model, defined by the joint probability distribution 
\begin{equation}
\label{Ising}
\Pr\left(s\right) = \frac{1}{\mathcal{Z}}\exp{\left[ \sum_{(ij)\in E} J_{ij}s_is_j + \sum_{i \in V}h_{i}s_{i}\right] }\, ,
\end{equation}
where $G = (V=[n],E)$ is an arbitrary undirected graph, with $n$ vertices and $m$ edges, $s\in\{\pm 1\}^{n}$ is a collection of binary spins, and the \emph{partition function} $\mathcal{Z}$ is defined by 
\begin{equation}
\label{partition_function}
\mathcal{Z} = \underset{s\in\{\pm 1\}^{n}}{\sum}\  \underset{(ij)\in E}{\prod}\ \exp{ (J_{ij}s_{i}s_{j})}\ \underset{i \in V}{\prod} \exp{(h_{i}s_{i}) } \, .
\end{equation}
We will call this model \emph{ferromagnetic} if the couplings $(J_{ij})_{(ij)\in E}$ and the fields $(h_{i})_{i\in[n]}$ are all positive. 
It can be shown (\cite{WJ2008}) that the couplings and the fields of the model \eqref{Ising} are in one-to-one correspondence with its \emph{magnetizations} $(m_{a})_{a\in[n]}$ and \emph{susceptibility matrix} $\chi\in\mathbb{R}^{n\times n}$, defined by
\begin{align}
 \label{magnetizations}
 m_{a} &= \E(s_{a}) = \underset{s\in\{\pm 1\}^{n}}{\sum} 
 s_{a}\ \Pr(s)\qquad&\text{ for }a\in[n]\, ,\\ 
 \label{susceptibility}
\chi_{ab} &= \E(s_{a}s_{b}) = \underset{s\in\{\pm 1\}^{n}}{\sum} s_{a}s_{b}\ \Pr(s)\qquad&\text{ for }a,b\in[n]\, .
 \end{align} 
The task of computing the quantities (\ref{partition_function},\ref{magnetizations},\ref{susceptibility}) knowing the joint probability distribution (\ref{Ising}) is sometimes referred to as \emph{inference}, or direct problem. Its applications include e.g. reconstructing partially observed binary images in computer vision \cite{greig1989exact}, or similarity-based clustering into $2$ groups \cite{blatt1996superparamagnetic}.
Conversely, the task of computing the joint probability distribution (\ref{Ising}) knowing the magnetizations (\ref{magnetizations}) and correlations (\ref{susceptibility}) is usually called \emph{learning}, or inverse problem. The practical importance of this inverse task stems from the fact that the model (\ref{Ising}) is the maximum entropy model with constrained means and correlations. In particular, it has found numerous applications in biology \cite{barton2016ace,morcos2011direct}, from predicting the three-dimensional folding of proteins, to identifying neural activity patterns. In machine learning, a variant of this inverse problem 
is usually referred to as Boltzmann Machine learning \cite{ackley1985learning}, and is a prototypical unsupervised learning problem. In practice, learning is done through a local optimization of the likelihood function, which involves the partition function (\ref{partition_function}), and which gradients involve the magnetizations (\ref{magnetizations}) and correlations (\ref{susceptibility}). 
 
Despite their considerable practical importance, the estimation of these three quantities in general is a notoriously intractable problem, except in particular cases, notably when the graph $G$ is planar and the external fields $(h_{i})_{i\in[n]}$ vanish. In the latter case, explicit expressions exist, originating in the work of Kac and Ward \cite{kac1952combinatorial,kasteleyn1963dimer,fisher1966dimer}, and allowing to devise polynomial-time inference algorithms \cite{schraudolph2009efficient}.
Recently, \cite{johnson2015learning} proposed a greedy algorithm to approximate an arbitrary graph by a planar one, thus making inference and learning tractable in a more general setting. Their approach, however, does not provide bounds on their estimates.     

For the case of the ferromagnetic Ising model (\ref{Ising}) with positive couplings $(J_{ij})_{(ij)\in E}$ and fields $(h_{i})_{i\in[n]}$, it is well known \cite{kolmogorov2004energy} that the ground-state, i.e. the configuration of the spins with minimum energy, can be found in polynomial time using graph cuts. This method has been used successfully in a number of computer vision problems, (see e.g. \cite{boykov2001fast,boykov1998markov,greig1989exact}), where a ferromagnetic Ising model is used to denoise a partially observed image. Under the same ferromagnetic assumption, sometimes called attractive or log-supermodular in the machine learning community, \cite{ruozzi2012bethe,willsky2008loop} showed that the stationary points of the Bethe free energy allow to lower bound the partition function (\ref{partition_function}). The proof of \cite{willsky2008loop} relies on a loop series expansion of the partition function first derived by \cite{chertkov2006loop}.
In this paper, we show how to \emph{upper} bound the marginals, correlations and partition function of the ferromagnetic Ising model. Interestingly, our results also have a strong connection with the Bethe approximation (see section \ref{sec:belief_prop_and_susceptibility_prop}).

The difficulty of inference and learning in the model (\ref{Ising}) has prompted the development of a wealth of numerical methods to approximate the quantities (\ref{partition_function},\ref{magnetizations},\ref{susceptibility}). In the ferromagnetic case, a particularly important breakthrough originated in the so-called cluster Monte Carlo methods of \cite{swendsen1987nonuniversal,wolff1989collective}. By updating a whole cluster of spins instead of just one, these algorithms can make non-local moves in the configuration space while still verifying the detailed balance condition. Therefore, they provide a substantial speed-up over conventional Markov Chain Monte Carlo methods, especially when the model (\ref{Ising}) is near criticality. Other numerical methods are typically based on a low or high temperature expansion of the quantities of interest, and an exhaustive enumeration of an increasing number of the terms contributing to this expansion \cite{campostrini200225th}. On the learning side of the problem, \cite{cocco2011adaptive} introduced a principled and accurate approximation scheme based on the identification of the clusters of spins that contribute the most to the entropy of the model. 
While potentially allowing to reach an arbitrary accuracy in the determination of the quantities (\ref{partition_function},\ref{magnetizations},\ref{susceptibility}), these numerical approaches do not, by nature, admit a closed-form solution, and do not in general provide bounds on their estimates.  

In this work, we prove simple and explicit upper bounds on the three quantities of interest (\ref{partition_function},\ref{magnetizations},\ref{susceptibility}) for \emph{arbitrary} graphs. These bounds are valid under two assumptions. First, we consider the particular case where the couplings $(J_{ij})_{(ij)\in E}$ and the fields $(h_i)_{i\in [n]}$ are 
positive. Second, we require the model (\ref{Ising}) to be in a well-defined high temperature region, specified by a condition on the spectral radius of a linear operator associated with the model (\ref{Ising}).  
Our results use the same starting point as some the intensive numerical methods described previously, but are much simpler in nature, and provide an efficient closed-form bound, valid on any finite graph, as long as the couplings and the fields are positive. We therefore expect our results to find natural applications, e.g. in the inference and learning examples listed above, where the computation of quantities such as the partition function (\ref{partition_function}) and the marginals (\ref{magnetizations},\ref{susceptibility}) play a central role. 

Our approach is based on the high temperature expansion of the Ising model, reviewed in section \ref{sec:high_T_expansion}. For ferromagnetic models, this expansion is composed of positive contributions corresponding to certain paths on the graph $G$. Our upper bounds are obtained by noting that the set of such paths is included in a tractable set of more general walks, for which an analytical expression can be derived. In this sense, our methodology is reminiscent of the one used in \cite{fisher1967critical} who derived similar upper bounds. The crucial difference between our work and \cite{fisher1967critical} is that whereas \cite{fisher1967critical} derives formulas for specific regular lattices in the thermodynamic limit, our results are applicable to arbitrary finite graphs and arbitrary positive couplings and are therefore not only of theoretical interest, but can be readily incorporated into data processing algorithms.  

\subsection{Notations and definitions}
We will denote the set of edges of a graph $G$ as $E(G)$. We will call $n$ the number of vertices in $G$ and $m$ the number of edges. For any integer $k \geq 1$, we denote by $[k]$ the set of integers greater or equal to $1$ and lower or equal to $k$. 
For any vertex $i\in [n]$, $\partial i$ will denote the set of neighbors of $i$ in the graph $G$.
Recall that the spectral radius $\rho(A)$ of a linear operator $A$ is defined as 
\begin{align}
\rho(A) = \max_{\lambda \in \text{Sp}(A)} \lvert \lambda\rvert\, ,
\end{align}
where Sp$(A)$ denotes the eigenvalue spectrum of $A$. 
For the Ising model in vanishing fields (\ref{Ising_zero_field}) on a finite graph $G$, we define the non-backtracking operator $\B \in \mathbb{R}^{2m \times 2m}$, acting on the directed edges of the graph, by its elements   
\begin{align}
\label{def:B}
\B_{(i\to j),(k\to l)} = \tanh( J_{kl})\ind(i = l)\ind(j \neq k)\, .
\end{align}
Finally, we define the Bethe Hessian operator $\H\in\mathbb{R}^{n\times n}$ with elements 
\begin{equation}
\label{def:H}
\H_{ij} = \ind(i=j)\left( 1 + \sum_{k\in\partial i}\frac{\tanh(J_{ik})^{2}}{1 - \tanh(J_{ik})^{2}} \right) - \ind(j\in\partial i)\frac{\tanh(J_{ij})}{1 - \tanh(J_{ij})^{2}}\, .
\end{equation}  
The non-backtracking operator, known as the Hashimoto matrix in graph theory \cite{hashimoto1989zeta}, was first introduced in the context of inference by \cite{krzakala2013spectral}, who used it as the basis of a spectral community detection method. Other applications of the non-backtracking operator to retrieval in the Hopfield model and to similarity-based clustering are studied in \cite{zhang2014non,saade2016clustering}. The use of the Bethe Hessian in inference was pioneered in \cite{saade2014spectral}, where its connection with the non-backtracking operator is highlighted in the context of community detection. Other applications to matrix completion and similarity-based clustering are considered in \cite{saade2015matrix,saade2016clustering}. Note that the Bethe Hessian corresponds to the Hessian of the Bethe free energy of the Ising model \eqref{Ising_zero_field}, i.e. to the inverse susceptibility matrix in the Bethe approximation \cite{saade2015matrix}.

\subsection{Assumptions}
\label{subsec:assumptions}
Our results hold for arbitrary finite graphs $G$, provided the two following conditions hold.
\begin{itemize} 
\item First, we assume that all the couplings $(J_{ij})_{(ij)\in E}$ as well as the external fields $(h_{i})_{i\in [n]}$ are non-negative. \item Second, we assume that the model (\ref{Ising}) is in a high temperature region, specified by the condition on the spectral radius of the non-backtracking matrix $\rho(\B) < 1$. 
\end{itemize}

It will prove convenient to restrict our analysis to the Ising model in vanishing external fields, defined by
\begin{equation}
\label{Ising_zero_field}
\P(s) = \frac{1}{\mathcal{Z}}\exp{\sum_{(ij)\in E} (J_{ij}s_is_j})\, .
\end{equation}
This is in fact not a restriction, as any $n$-spin Ising model with fields can be expressed as a $n+1$-spin model without fields, so that model (\ref{Ising_zero_field}) is as expressive as model (\ref{Ising}) \cite{fisher1967critical}. To make this claim precise, we will use proposition 1 in \cite{johnson2015learning}, which we recall here for completeness.

\begin{proposition}
\label{reduction_no_fields}
(proposition 1 in \cite{johnson2015learning}) Consider the Ising model (\ref{Ising}) on the graph $G=([n],E(G))$, with couplings $(J_{ij})_{(ij)\in E(G)}$ and fields $(h_{i})_{i\in[n]}$, and corresponding partition function $\mathcal{Z}$, magnetizations $(m_{a})_{a\in [n]}$, and correlations $(\chi_{ab})_{(a,b)\in[n]^{2}}$. Define another Ising model on the graph $\hat{G} = ([n+1],E(G)\cup\{(i,n+1),\forall i\in[n]\})$ with vanishing fields, and couplings 
\begin{equation}
\hat{J}_{ij} = 
\left\{
\begin{split}
J_{ij} \text{ if } j < n+1 \\ 
h_i \text{ if } j = n+1
\end{split}
\right.\, .
\end{equation}
Call $\hat{\mathcal{Z}}$ its partition function, and $(\hat{\chi}_{ab})_{(a,b)\in[n+1]^{2}}$ its correlations. Then $\hat{\mathcal{Z}} = 2\mathcal{Z}$ and 
\begin{equation}
\hat{\chi}_{ab} = 
\left\{
\begin{split}
\chi_{ab} \text{ if } a < n+1 \text{ and } b < n+1 \\ 
m_a \text{ if } a < n+1 \text{ and } b = n+1
\end{split}
\right.\, .
\end{equation}
\end{proposition}

We will therefore consider model (\ref{Ising_zero_field}) in the following, and bound its partition function and correlations, which will yield a bound on the partition function, magnetizations and correlations of model (\ref{Ising}). Note also that from proposition \ref{reduction_no_fields}, the new couplings $(\hat{J}_{ij})_{(ij)\in E(\hat{G})}$ are positive if and only if both the original couplings $({J}_{ij})_{(ij)\in E(G)}$ and fields $(h_{i})_{i\in[n]}$ are positive.

\subsection{Algorithms}

Before stating our rigorous results in the next section, we describe here the corresponding procedures to upper bound the partition function, magnetizations and correlations of the Ising model on a given graph. As will be apparent from the upcoming results, our bounds can be expressed either using the non-backtracking operator or the Bethe Hessian. Compared to the former, the latter is a smaller, real and symmetric matrix. We therefore present here algorithms relying on the Bethe Hessian in the interest of numerical efficiency. 
 
As stated in the previous section, when given a general Ising model  with finite fields of the form \eqref{Ising}, our first step, common to both the two upcoming procedures, is to define an associated Ising model in vanishing fields of the form \eqref{Ising_zero_field}. By bounding the partition function and correlations of the resulting model in vanishing fields, we obtain bounds on the partition function, magnetizations, and correlations of the original model using the correspondence recalled in proposition \ref{reduction_no_fields}.

\begin{algorithm}
\caption{Bound on the log partition function of the Ising model in vanishing fields}
\label{alg:bound_partition}
\begin{algorithmic}[1]
\Require{Graph $G = (n,E(G))$, couplings $(J_{ij})_{(ij)\in E(G)}$}
\Ensure{Upper bound $\mathcal{L}$ on the log partition function $\log\mathcal{Z}$}
\State Build the Bethe Hessian of equation \eqref{def:H}
\State Output $\mathcal{L} =n\log 2 - \frac{1}{2}\log\det (\H) + 2\sum_{(ij)\in E(G)} \log\cosh(J_{ij})$
\end{algorithmic}
\end{algorithm}

Algorithm \ref{alg:bound_partition} describes our procedure to estimate the logarithm of the partition function. We will show in the following that under the assumptions of section \ref{subsec:assumptions}, this simple algorithm yields an upper bound that we expect to be most accurate on graphs containing few (or large) loops. We note that on sparse graphs, the logarithm of the determinant of the Bethe Hessian can be computed efficiently by first performing a Cholesky decomposition of the sparse matrix $\H$. 
\begin{algorithm}
\caption{Bounds on the correlations of the Ising model in vanishing fields}
\label{alg:bound_correlations}
\begin{algorithmic}[1]
\Require{Graph $G = (n,E(G))$, couplings $(J_{ij})_{(ij)\in E(G)}$}
\Ensure{Upper bounds $c_{ab}$ on the correlations $\chi_{ab}$, for $a,b\in[n]$}
\State Build the Bethe Hessian of equation \eqref{def:H}
\State Output $c_{ab} = \left(\H^{-1}\right)_{ab}$ for $a,b\in [n]$
\end{algorithmic}
\end{algorithm}

Algorithm \ref{alg:bound_correlations} yields an estimate of the correlations in the Ising model with vanishing fields which we will show to be an upper bound under the conditions of section \ref{subsec:assumptions}. Note again that this algorithm can be used to compute an upper bound on the magnetizations of an Ising model with finite fields using proposition \ref{reduction_no_fields}. Once more, for sparse models, this procedure can be made numerically efficient by using sparse linear solvers instead of directly inverting the Bethe Hessian.

\section{Main results}
We now state our main results and some of their consequences, leaving the proofs to the next section. 
\subsection{Bound on the partition function}
\label{sec:Bound_on_partition_function}
Our first result is an upper bound on the partition function (\ref{partition_function}).
\begin{theorem}
\label{th:bound_Z}
Consider model (\ref{Ising_zero_field}) with positive couplings $J_{ij} > 0$ for $(ij)\in E(G)$. Assume that the spectral radius $\rho(\B) <1$, where $\B$ is the non-backtracking matrix defined in (\ref{def:B}), and let $\H$ be the Bethe Hessian defined in (\ref{def:H}). Then 
\begin{equation}
\label{bound_Z}
\mathcal{Z} \leq 2^{n}\det\left(\I - \B\right)^{-1/2}\prod_{(ij)\in E(G)} \cosh(J_{ij})
= 2^{n}\det (\H)^{-1/2}\prod_{(ij)\in E(G)} \cosh(J_{ij})^{2}\, .
\end{equation}
\end{theorem}
Note that the non-backtracking matrix is a large $(2m\times 2m)$, non-symmetric object, non-trivial to build or manipulate. The equality in (\ref{bound_Z}) allows to compute this bound without having to build the non-backtracking matrix, by using instead the smaller ($n\times n$) and symmetric Bethe Hessian. This last equality is a simple consequence of the Ihara-Bass formula, which, in its multivariate version (\cite{watanabe2009graph}, theorem 2) states that 
\begin{align}
\label{Ihara-Bass}
\det(\I - \B) = \det(\H)\prod_{(ij)\in E(G)}\cosh(J_{ij})^{-2}\, .
\end{align}
In particular, this formula implies that on any \emph{finite} graph, $\H$ is non-singular if $\rho(\B)<1$. In the thermodynamic limit $n\to\infty$, some subtleties arise, as discussed in the next section.

\par
\begin{figure}[H]
\begin{center}
  \includegraphics[width=1.0\textwidth]{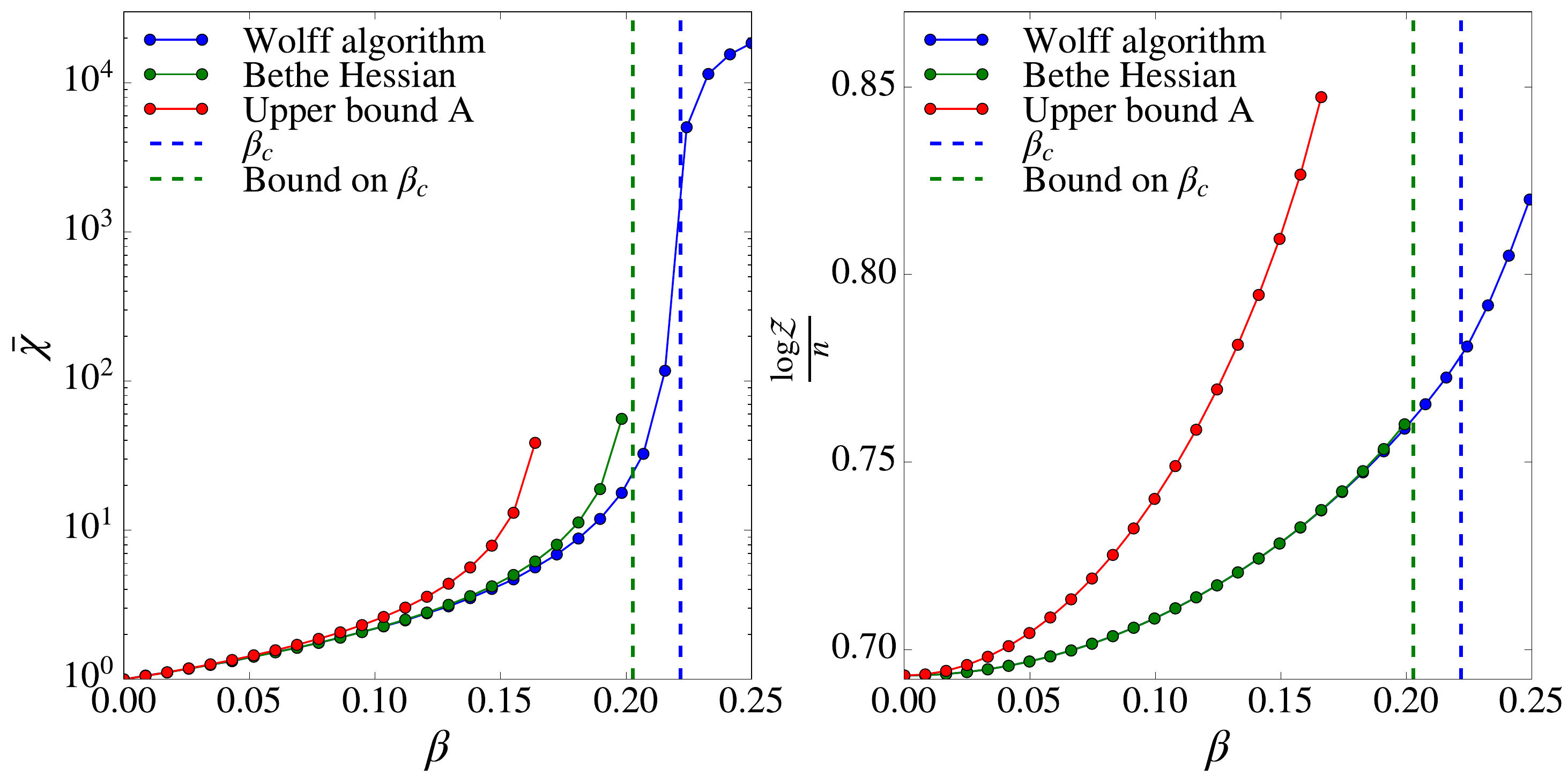}
\caption{Numerical simulation of the 3D lattice Ising model with $n=32^3$ spins, with periodic boundary conditions, and uniform couplings $J_{ij} = \beta$.  
The left panel is the scalar susceptibility $\bar{\chi}$ of equation (\ref{scalar_susceptibility}), and the right panel is the logarithm of the partition function defined in equation (\ref{partition_function}). For both quantities we represented the exact value obtained via a computationally expensive Monte Carlo simulation (using the Wolff algorithm \cite{wolff1989collective}), and the upper bounds (\ref{bound_Z}) and (\ref{bound_susceptibility_with_H}) expressed in terms of the Bethe Hessian operator. The dashed lines signal the critical temperature $\beta_c$ of the 3D Ising model as computed numerically in \cite{preis2009gpu}, and the lower bound on this critical temperature provided by the Bethe Hessian (\ref{bound_beta_c_regular}). Finally, we included for comparison the upper bound obtained by using the adjacency matrix $A$ instead of the non-backtracking matrix (equations (\ref{bound_Z_with_A}) and (\ref{bound_susceptibility_with_A})).
}
\label{fig:3Dising}       
\end{center}
\end{figure}

 As will be apparent from the proof, this bound is based on an over-counting of the subgraphs of $G$ contributing to the high temperature expansion of the partition function (see section \ref{sec:high_T_expansion}). These subgraphs correspond to closed paths (i.e. closed self-avoiding walks) that are notoriously hard to count \cite{fisher1964ising}. To derive an analytical upper bound on the partition function, the set of contributing subgraphs must be included in a larger set of walks on the graph $G$ which contribution can be computed analytically. 
A simple bound can be obtained by including the set of closed paths in the set of all closed walks on the graph $G$. This approach yields an analytical bound similar to Theorem \ref{th:bound_Z} expressed in terms of the adjacency matrix of the graph, namely 
\begin{equation}
\label{bound_Z_with_A}
 \mathcal{Z} \leq 2^{n}{\det\left(\I - A\right)}^{-1}\prod_{(ij)\in E(G)} \cosh(J_{ij})  \, ,
 \end{equation} 
whenever $\rho(A)< 1$, where $A$ is the weighted adjacency matrix of the graph $G$, with entries $A_{ij} = \tanh(J_{ij})\ind(i\in \partial j)$ (see section \ref{sec:proof_theo_Z} for details). This bound is, however, too loose, because it counts many spurious contributions, in particular walks that are allowed to go back and forth on the same edge. We improve this bound by forbidding that the walk immediately backtracks to the previous edge. This is achieved by replacing the adjacency matrix with the non-backtracking operator. From the proof of section \ref{sec:proof_theo_Z}, it is straightforward to see that the bound (\ref{bound_Z}) is less tight if the graph $G$ contains many loops. 
In practice, however, figure \ref{fig:3Dising} shows that using the non-backtracking operator instead of the adjacency matrix leads to a dramatic improvement, even in the case of a 3D lattice Ising model, which has many loops.    


\subsection{Bound on the susceptibility}
We now state our upper bound on the correlations, encoded in the susceptibility matrix $\chi$ of equation~(\ref{susceptibility}).
\begin{theorem}
\label{th:bound_susceptibility_with_B}
Consider model (\ref{Ising_zero_field}) with positive couplings $J_{ij} > 0$ for $(ij)\in E(G)$. Assume that $\rho(\B) <1$, where $\B$ is the non-backtracking matrix defined in (\ref{def:B}). Define the matrices $\P,\Q\in\mathbb{R}^{n\times 2m}$ by their elements, for $a\in [n],(ij)\in E(G)$:
\begin{equation}
\label{P&Q}
\P_{a,(i\to j)} = \tanh(J_{ij})\ind(a = j),\qquad \Q_{a,(i\to j)} = \ind(a = i) \, .
\end{equation}
Then
\begin{equation}
\label{bound_susceptibility_with_B}
\chi \leq \P(\I - \B)^{-1}\trans{\Q} + \I_{n\times n}\, ,
\end{equation}
where the inequality holds element-wise.
\end{theorem}
Once more, it is possible to express this last result in terms of the Bethe Hessian rather than the non-backtracking operator, yielding a surprisingly simple result.

\begin{corollary}
\label{cor:bound_susceptibility_with_H}
Let $\H$ be the matrix defined in (\ref{def:H}). Under the same assumptions as theorem \ref{th:bound_susceptibility_with_B}, it holds that $\H$ is invertible, and 
\begin{equation}
\label{bound_susceptibility_with_H}
\chi \leq \H^{-1}\, ,
\end{equation}
where the inequality holds element-wise.
\end{corollary} 

\par As for the partition function, these bounds rely on the inclusion of the set of paths between two spins in a larger, tractable set of walks on the graph $G$ (see section \ref{sec:proof_susceptibility_B}). More precisely, expression (\ref{bound_susceptibility_with_B}) follows from the inclusion of the set of paths between two fixed spins $a,b\in[n]$ in the set of non-backtracking walks starting at $a$ and ending at $b$. If we allow these walks to backtrack, i.e. if we include the set of paths in the set of all walks, we get the looser bound 
\begin{equation}
\label{bound_susceptibility_with_A}
 \chi \leq (\I_{n} - A)^{-1}\, ,
 \end{equation} 
whenever $\rho(A) < 1$ where $A$ is the weighted adjacency matrix with elements $A_{ij} = \tanh(J_{ij})\ind(i\in \partial j)$. Once more, forbidding backtracking allows to dramatically improve the bound on the susceptibility, as shown in figure \ref{fig:3Dising}.

\par While valid only on finite graphs, these results allow us, in certain cases, to derive a bound on the paramagnetic to ferromagnetic phase transition. 
More precisely, for a sequence of Ising models, the previous results allow to bound the scalar susceptibility, defined as 
\begin{equation}
\label{scalar_susceptibility}
\bar{\chi} := \frac{1}{n} \sum_{x,y = 1}^{n} \chi_{xy}\, .
\end{equation}

\begin{corollary}
\label{cor:bound_phase_transition}
Consider a sequence $(I_{p})_{p\in \mathbb{N}}$ of Ising models of the form (\ref{Ising_zero_field}), each of them defined on a graph $G_{p}$, with positive couplings $J^{(p)}_{ij}>0$, for $(ij)\in E(G_{p})$, and scalar susceptibility (\ref{scalar_susceptibility}) denoted $\bar{\chi}_{p}$. Define $\B_{p}$ to be the non-backtracking operator defined in (\ref{def:B}) for the Ising model $I_{p}$, and assume that $\forall p\in\mathbb{N}$, $\rho(\B_{p})<1$.
Define $\H_{p}$ to be the Bethe Hessian defined in (\ref{def:H}) for the Ising model $I_{p}$, and let $\lambda_{\min}(\H_{p})$ denote its smallest eigenvalue. 
Assume that there exists $\epsilon > 0$ such that $\forall p\in\mathbb{N}$, $\lambda_{\min}(\H_{p})>\epsilon$. Then $\forall p\in\mathbb{N}$,
\begin{equation}
\bar{\chi}_{p} \leq \frac{1}{\epsilon}
\end{equation}
\end{corollary} 

\noindent\emph{Proof:}
For any fixed $p\in\mathbb{N}$, let $n_{p}$ be the number of vertices of the graph $G_{p}$.
Defining $\U_{p}\in\mathbb{R}^{n_{p}}$ to be the vector with all its entries equal to $1$, and denoting by $||.||_{p}$ the Euclidean norm, we have from Corollary \ref{cor:bound_susceptibility_with_H}
\begin{equation}
\bar{\chi}_{p} \leq \frac{1}{n_{p}} \sum_{x,y = 1}^{n_{p}} (\H_{p})^{-1}_{xy} = \frac{\trans{\U_{p}}\H_{p}^{-1}\U_{p}}{||\U_{p}||_{p}^{2} }\leq \rho(\H_{p}^{-1}) = \frac{1}{\lambda_{\min}(\H_{p})}\leq\frac{1}{\epsilon}\, ,
\end{equation}
where we have used that the matrix $\H_{p}$ is symmetric. 
\medskip

\par For sequences of graphs that admit a thermodynamic limit, Corollary \ref{cor:bound_phase_transition} implies a condition under which the infinite Ising model is in the paramagnetic phase, therefore yielding a bound on the paramagnetic to ferromagnetic transition. As an explicit example, let us consider the case of a sequence $(G_{p})_{p\in\mathbb{N}}$ of $d$-regular graphs with uniform couplings $J_{ij}=\beta$. We assume that the number of vertices in $G_{p}$ goes to infinity as $p\to\infty$. It is straightforward to check that, for any $p\in\mathbb{N}$,
\begin{equation}
\rho(\B_p) = (d-1)\tanh(\beta)\qquad\text{ and }\qquad  \lambda_{\min}(\H_{p}) = 1 - \frac{d\tanh(\beta)}{1+\tanh(\beta)}\, .
\end{equation}
By application of Corollary \ref{cor:bound_phase_transition}, it follows that if
\begin{equation}
(d-1)\tanh(\beta) < 1\, ,
\end{equation}
the scalar susceptibility $\bar{\chi}_{p}$ remains bounded. Equivalently, if the sequence admits a thermodynamic limit with a phase transition at $\beta_{c}$, then 
\begin{equation}
\label{bound_beta_c_regular}
\beta_{c}\geq \atanh\frac{1}{d-1}\, .
\end{equation}
Interestingly, the right hand side of the last equation corresponds to the critical inverse temperature in the Bethe approximation, which is already known to be a lower bound on $\beta_{c}$ for certain ferromagnetic models on hypercubical lattices \cite{fisher1967critical}. Our contribution generalizes such previous results, giving a simple algorithm to derive a lower bound on arbitrary graphs, with arbitrary (positive) couplings.  
It is worth noting that, perhaps unsurprisingly, this bound is tight on sparse random $d$-regular graphs with uniform couplings $\forall (ij)\in E(G), J_{ij} = \beta$. Indeed, on such locally tree-like graphs, the existence of the thermodynamic limit has been proved, and, the critical temperature has been shown to verify $(d-1) \tanh (\beta) = 1$ \cite{dembo2010ising,dommers2014ising}. More generally, 
we expect the implicit bound provided by Corollary \ref{cor:bound_phase_transition} on the transition of the Ising model to be tight on any sparse random graph ensemble whose degree distribution has a finite second moment.

\par From the Ihara-Bass formula (\ref{Ihara-Bass}), it holds that on any \emph{finite} graph, $\lambda_{\min}(\H) >0$ as long as $\rho(\B) < 1$. Therefore one may expect that the condition on the smallest eigenvalue of $\H_{p}$ in Corollary \ref{cor:bound_phase_transition} could be relaxed, e.g. by assuming instead that $\rho(\B) < 1-\epsilon$ for some $\epsilon > 0$. This relaxation turns out to be not possible, because it may happen that the spectral radius of $\B$ is bounded away from $1$, while the smallest eigenvalue of $\H$ tends to $0$, the limit $p\to\infty$. As an example, take $G_{p}$ to be the star graph with $n_{p} = p+1$ spins and edges $(i,n+1)$ for $i\in [n]$, and uniform couplings $J_{ij}=\beta$. Since $G_{p}$ is a tree, and the non-backtracking operator is nilpotent on trees, it holds that $\rho(\B_{p}) = 0$ for all $p$. On the other hand, the spectrum of $\H_{p}$ can be computed explicitly, and its smallest eigenvalue shown to verify 
\begin{equation}
\lambda_{\min}(\H_{p}) \underset{p\rightarrow\infty}{\sim} \frac{1}{p\tanh(\beta)^{2}} \underset{p\rightarrow\infty}{\longrightarrow}\, .
0\end{equation}

\subsection{Relation to belief propagation and susceptibility propagation}
\label{sec:belief_prop_and_susceptibility_prop}
A standard approximation to compute the magnetizations of model (\ref{Ising_zero_field}) is the belief propagation algorithm, or cavity method in statistical physics \cite{mezard2009information}, which solutions can be shown to yield stationary points of the Bethe free energy \cite{yedidia2001bethe}. Belief propagation is a recursive message-passing algorithm that approximates the magnetizations $m_{i} = \E(s_{i})$ as 
\begin{equation}
\label{BP_full}
m_i \approx \tanh\left(\sum_{l\in \partial i} \atanh\left( m_{l\to i} \tanh{J_{il}} \right)  \right)\, ,
\end{equation}
where the so-called cavity fields $(m_{i\to j})_{(i\to j)\in \vec{E}(G)}$ are defined on the set $\vec{E}(G)$ of directed edges of $G$, and verify the fixed point equation 
\begin{equation}
\label{BP_recursion}
m_{i\to j} = \tanh\left(\sum_{l\in \partial i\backslash j} \atanh\left( m_{l\to i} \tanh{J_{il}} \right)  \right)\, .
\end{equation}
In practice, starting from a random initial condition, one iterates equation (\ref{BP_recursion}) until convergence, and outputs the result of (\ref{BP_full}). 
Belief propagation is known to be exact on trees, and widely believed to yield asymptotically accurate results for sparse, locally-tree like graphs,
as well as models with small couplings \cite{mezard2009information}.  

The message-passing approach can be extended to allow the computation of the correlations $\chi_{ij}$ for any $i,j\in[n]$ using the fluctuation dissipation theorem. The resulting algorithm is called susceptibility propagation \cite{mora2007geometrie,mezard2009constraint}, and approximates $\chi_{ij}$ as 
\begin{equation}
\label{SP_full}
\chi_{ij} \approx (1 - m_{i}^{2})\left( \ind(i =j) + \sum_{l\in\partial i} \frac{\chi_{l \to i, j} \tanh{J_{il}}}{1 - m_{l\to i}^{2}\tanh^{2}J_{il}}    \right)\, ,
\end{equation}
where the $( m_{i\to j})_{(i\to j)\in \vec{E}(G)}$ are the fixed point of (\ref{BP_recursion}), the $(m_{i})_{i\in[n]}$ are the belief propagation estimates of the magnetization derived from (\ref{BP_full}) and the $(\chi_{i\to j,k})_{(i\to j)\in \vec{E}(G),k\in [n]}$ verify the fixed point equation 
\begin{equation}
\label{SP_recursion}
\chi_{i \to j,k} = (1 - m_{i\to j}^{2})\left( \ind(i = k) + \sum_{l\in\partial i\backslash j} \frac{\chi_{l \to i, k} \tanh{J_{il}}}{1 - m_{l\to i}^{2}\tanh^{2}J_{il}}    \right)\, ,
\end{equation}
Similarly to belief propagation, starting from a random initial condition, equation (\ref{SP_recursion}) is first iterated until convergence, and the correlations are then estimated from equation (\ref{SP_full}).
Note that it is possible to invert the relation between the susceptibilities and the couplings, resulting in an inference algorithm for solving the inverse Ising model. This algorithm has been shown to yield better results than other mean field approaches on certain problems (\cite{mora2007geometrie,ricci2012bethe}).     

While only exact on trees, both belief propagation and susceptibility propagation have been observed to yield good approximate results on more general topologies, when they converge. However, these algorithms are based on the Bethe approximation \cite{yedidia2001bethe}, which, unlike the naive mean-field approach, does not provide bounds on the actual partition function \cite{WJ2008}. The following result might therefore appear surprising.  

\begin{corollary}
Consider model (\ref{Ising_zero_field}) with positive couplings $J_{ij} > 0$ for $(ij)\in E(G)$. Assume that $\rho(\B) <1$, where $\B$ is the non-backtracking matrix defined in (\ref{def:B}). Then $(m_{i\to j} = 0)_{(i\to j)\in \vec{E}(G)}$ is a \emph{stable} fixed point of the belief propagation recursion (\ref{BP_recursion}). Additionally, the corresponding susceptibility propagation algorithm converges, and yields an upper bound on the true correlations, regardless of the topology of the graph.   
\end{corollary} 

\noindent\emph{Proof:}
The fact that $(m_{i\to j} = 0)_{(i\to j)\in \vec{E}(G)}$ is a fixed point of belief propagation is readily checked on equation (\ref{BP_recursion}). To see that it is stable, starting from a small perturbation $\delta m^{0}_{i\to j}$, we have at first order at iteration $t\geq1$
\begin{equation}
\delta m^{t}_{i\to j} = \sum_{l\in\partial i\backslash j} \delta m^{t-1}_{l\to i}\tanh{J_{il}}\, , 
\end{equation}
which in matrix form can be written $\delta m^{t} = \B\, \delta m^{t-1}$. The stability of this fixed point follows from the assumption $\rho(\B) < 1$. The corresponding belief propagation solution is $m_{i} = 0,\forall i\in [n]$.
The corresponding susceptibility propagation recursion reads 
\begin{equation}
\label{para_susceptibility_prop}
\chi_{i \to j,k} = \ind(i = k) + \sum_{l\in\partial i\backslash j} \chi_{l \to i, k} \tanh{J_{il}}\, .
\end{equation}
Define a vector $\chi_{k}\in\mathbb{R}^{2m}$ with elements $\chi_{i\to j,k}$, for $(i\to j)\in \vec{E}(G)$, and call $\Q_{k}$ the $k$-th line of the matrix $\Q$ defined in Theorem \ref{th:bound_susceptibility_with_B}. Then equation (\ref{para_susceptibility_prop}) can be rewritten in matrix form as 
\begin{equation}
\chi_{k} = \trans{\Q_{k}} + \B \chi_{k}\, .
\end{equation}
which solution $\chi_{k} = (\I - \B)^{-1}\trans{\Q_{k}}$ exists and is unique, since $\rho(\B) <1$. Iterating equation (\ref{para_susceptibility_prop}) starting from the initial condition $\chi^{0}_{k}$, we get at iteration $t \geq 1$
\begin{equation}
\chi^{t}_{k} - \chi_{k} = \B\left( \chi^{t-1}_{k} - \chi_{k} \right)\, ,
\end{equation}
so that $\chi^{t}_{k} \rightarrow \chi_{k}$ as $t\rightarrow \infty$, using again that $\rho(B)<1$. Finally, using equation (\ref{SP_full}), it is straightforward to check that the correlations output by susceptibility propagation are given in matrix form by $\P(\I - \B)^{-1}\trans{\Q} + \I_{n\times n}$, which is, by Theorem \ref{th:bound_susceptibility_with_B}, an upper bound on the true correlations. 

\section{Proofs}

\subsection{High temperature expansion}
\label{sec:high_T_expansion}
Central to our results is the high temperature expansion of the partition function, which we quickly rederive here. We use the following rewriting of the Boltzmann weight, relying on the fact that the spins are binary variables equal to $\pm 1$
\begin{equation}
\exp{( J_{ij}s_{i}s_{j} )} = a_{ij}(1 + b_{ij}s_{i}s_{j})\, ,
\end{equation}
with $a_{ij} = \cosh( J_{ij}), b_{ij} = \tanh( J_{ij})$. The partition function of model (\ref{Ising_zero_field}) is given by 
\begin{align}
\mathcal{Z} &= \underset{s\in\{\pm 1\}^{n}}{\sum}\ \  \underset{(ij)\in E(G)}{\prod}\ \exp{( J_{ij}s_{i}s_{j})}\\ 
& = \left(\prod_{(ij)\in E(G)} a_{ij}\right) \underset{s\in\{\pm 1\}^{n}}{\sum}\ \  \underset{(ij)\in E(G)}{\prod}\ (1 + b_{ij}s_{i}s_{j}) \\ 
& = \left(\prod_{(ij)\in E(G)} a_{ij}\right) \underset{s\in\{\pm 1\}^{n}}{\sum} \left[1+ \underset{(ij)\in E(G)}{\sum}\ b_{ij}s_{i}s_{j} + \underset{(ij),(kl)\in E(G)}{\sum}\ b_{ij}b_{kl}s_{i}s_{j}s_{k}s_{l} + \cdots\right]\, .
\end{align}
After summing on the spin configurations $s\in\{\pm 1\}^{n}$, the only terms in the expansion that do not vanish are the ones supported on a subgraph of $G$ where all nodes have even degree. These are closed paths (not necessarily connected). Each of these closed paths contribute a factor $2^{n}$ to the partition function. We can therefore rewrite the partition function in the following form, called \emph{high temperature expansion}
\begin{equation}
\label{highTexpansion}
\mathcal{Z} = 2^{n}\left(\prod_{(ij)\in E(G)} a_{ij}\right)\left[1 + \underset{g\in \mathcal{C}}{\sum} \prod_{(ij)\in E(g)} b_{ij} \right]\, ,
\end{equation}
where ${\cal C}$ is the set of closed paths, possibly disconnected.

\subsection{Proof of Theorem \ref{th:bound_Z}}
\label{sec:proof_theo_Z}
We now introduce the set $\mathcal{C}_{l}^{c}$ of \emph{connected} closed paths of length $l$. 
We denote by ${\cal W}_{l}^{c}$ the sum of all contributions to the partition function coming from connected closed paths of length $l\geq 1$, i.e.
\begin{align}
{\cal W}_{l}^{c} = \underset{g\in \mathcal{C}_{l}^{c}}{\sum}\ \prod_{(ij)\in E(g)} b_{ij}\, .
\end{align}
We have the inequality
\begin{align}
\mathcal{Z} &\leq 2^{n}\left(\prod_{(ij)\in E(G)} a_{ij}\right)\left[1 + \sum_{l\geq1}\ \   \sum_{n_c\geq1}\ \  \frac{1}{n_{c}!} \sum_{l_{1} + l_{2} + \cdots + l_{n_{c}} = l} {\cal W}_{l_{1}}^{c} {\cal W}_{l_{2}}^{c} \cdots {\cal W}_{l_{n_{c}}}^{c}\right] \\
& \leq 2^{n}\left(\prod_{(ij)\in E(G)} a_{ij}\right)\left[1 +  \sum_{n_c\geq1}\ \  \frac{1}{n_{c}!} \left(\sum_{l\geq1} {\cal W}_{l}^{c} \right)^{n_{c}}\right]\, .
\label{pause_bounding}
\end{align}
Indeed, the product ${\cal W}_{l_{1}}^{c} {\cal W}_{l_{2}}^{c} \cdots {\cal W}_{l_{n_{c}}}^{c}$ contains all the contributions coming from disconnected closed paths which connected components are of size $l_{1},\cdots,l_{n_{c}}$. The factor $n_{c}!$ accounts for all the permutations of the factors in this product. This is only an inequality because we are counting graphs in which some edges appear more than once. However, the factors ${\cal W}_{l}^{c}$ are still hard to compute, and we will look for an upper bound. A simple upper bound can be derived by considering a weighted version of the adjacency matrix of the graph. Define $A\in\mathbb{R}^{n\times n}$ by its entries
\begin{align}
A_{ij} = \left\{
    \begin{array}{ll}
        b_{ij} & \text{ if } (ij)\in E(G) \\
        0 & \text{ otherwise}
    \end{array}
\right.\, ,
\end{align}
then it holds that 
\begin{align}
{\cal W}_{l}^{c} \leq \frac{\Tr{A^{l}}}{l}\, .
\end{align}
Indeed, the right hand side contains all contributions coming from closed walks that start and end at the same point. The factor $1/l$ accounts for the choice of the starting point. However, as hinted at in section \ref{sec:Bound_on_partition_function}, this upper bound is loose, because the adjacency matrix allows backtracking, and therefore going back and forth on the same edge, whereas such contributions do not appear in the partition sum. To improve this bound, we use a matrix that forbids backtracking: the operator $\B$ of equation (\ref{def:B}). 
Since connected closed paths are closed non-backtracking walks (although the converse is not true), it holds that  
\begin{align}
{\cal W}_{l}^{c} \leq \frac{\Tr{\B^{l}}}{2l}\, .
\end{align}
where the additional factor $1/2$ accounts for the degeneracy due to the orientation of the non-backtracking closed walk. Under the assumption that $\rho(\B)<1$, the following bound holds:
\begin{align}
\sum_{l\geq1}{\cal W}_{l}^{c} \leq \sum_{l\geq1}\frac{\Tr{\B^{l}}}{2l} = - \frac{1}{2}\Tr \log(\I - \B) = -\frac{1}{2} \log \det(\I - \B)\, ,
\label{series}
\end{align}
where $\I\in\mathbb{R}^{2m\times 2m}$ is the identity matrix. This, along with the Ihara-Bass formula (\ref{Ihara-Bass}) completes the proof of Theorem \ref{th:bound_Z}.

\subsection{Proof of Theorem \ref{th:bound_susceptibility_with_B}}
\label{sec:proof_susceptibility_B}
The correlation functions are given, for any $x,y\in [n]$, by 
\begin{align}
\E(s_{x}s_{y})  &= \frac{\underset{s\in\{\pm 1\}^{n}}{\sum} s_{x}s_{y}  \underset{(ij)\in E(G)}{\prod}\ (1 + b_{ij}s_{i}s_{j})}{\underset{s\in\{\pm 1\}^{n}}{\sum}\ \  \underset{(ij)\in E(G)}{\prod}\ (1 + b_{ij}s_{i}s_{j})} =
 \frac{\underset{{g\in\mathcal{P}_{xy}+\mathcal{C}}}{\sum}\   \underset{(ij)\in E(g)}{\prod} b_{ij}           }{1+\underset{{g^{\prime}\in\mathcal{C}}}{\sum}\   \underset{(ij)\in E(g^{\prime})}{\prod} b_{ij}   } 
\end{align}
where: 
\begin{itemize}
\item $\mathcal{P}_{xy}$ is the set of paths from $x$ to $y$.
\item $\mathcal{C}$ is, as in the previous section, the set of closed paths.
\item $\mathcal{P}_{xy}+\mathcal{C}$ is the set of diagrams made of a path from $x$ to $y$ and any number of disconnected closed paths such that each edge is selected at most once. The closed paths therefore do not intersect the path from $x$ to $y$.
\end{itemize}
We assume without loss of generality that $x\neq y$. From the previous definitions, we have that
\begin{equation}
\E(s_{x}s_{y}) \leq \frac{\left(\underset{g\in\mathcal{P}_{xy}}{\sum}\   \underset{(ij)\in E(g)}{\prod} b_{ij} \right) \left( 1+\underset{{g^{\prime}\in\mathcal{C}}}{\sum}\   \underset{(ij) \in E(g^{\prime}) }{\prod} b_{ij} \right)}   {1+\underset{{g^{\prime}\in\mathcal{C}}}{\sum}\   \underset{(ij)\in E(g^{\prime})}{\prod} b_{ij}   } = \underset{g\in\mathcal{P}_{xy}}{\sum}\   \underset{(ij)\in E(g)}{\prod} b_{ij}\, .
\end{equation}
Indeed, developing the product at the numerator gives a sum of positive contributions including the ones in $\mathcal{P}_{xy}+\mathcal{C}$ and also (positive) spurious contributions coming from diagrams where the closed loops intersect the path from $x$ to $y$. We now introduce the set $\mathcal{N}_{(x \to x^{\prime}),(y^{\prime} \to y)}^{l}$ of non-backtracking walks of length $l$, starting on the directed edge $(x\to x^{\prime})$ and terminating on the edge $(y^{\prime} \to y)$. For a non-backtracking walk $w$, we denote by $\mathcal{E}(w)$ the list of edges crossed by $w$, where each edge appears with a multiplicity equal to the number of times $w$ crosses this edge.
Since any path in $\mathcal{P}_{xy}$ is a non-backtracking walk (though the reverse is again, in the presence of loops, not true), it holds that 
\begin{align}
\E(s_{x}s_{y}) \leq \sum_{l\geq 1}\ \  \sum_{\substack{x^{\prime} \in \partial x \\ y^{\prime} \in \partial y }}\ \  \sum_{w\in \mathcal{N}_{(x \to x^{\prime}),(y^{\prime} \to y)}^{l} }
\ \ \prod_{(ij)\in \mathcal{E}(w)} b_{ij}
\end{align} 
where $\partial x$ denotes the set of neighbors of $x$ in the graph $G$. In order to write this last expression in terms of the non-backtracking operator, we introduce the vector $u_{x \to x^{\prime}} \in\mathbb{R}^{2m}$ with entries all equal to $0$ except for the $(x \to x^{\prime})$ entry, which is equal to $1$. Similarly, we introduce the vector $v_{y^{\prime} \to y} \in\mathbb{R}^{2m}$ which only non-zero entry is equal to $b_{yy^{\prime}}$, in position $(y \to y^{\prime})$. Then we have 
\begin{align}
\sum_{g\in \mathcal{N}_{(x \to x^{\prime}),(y^{\prime} \to y)}^{l} } 
\ \ \prod_{(ij)\in E(g)} b_{ij} = \trans{(v_{y^{\prime} \to y})}\, \B^{l-1}\, u_{x\to x^{\prime}}\, ,
\end{align}
so that 
\begin{align}
\E(s_{x} s_{y}) &\leq  \trans{\left(\sum_{y^{\prime}\in\partial y}v_{y^{\prime} \to y}\right)} \sum_{l\geq1} \B^{l-1} \left(\sum_{x^{\prime}\in\partial x}u_{x \to x^{\prime}}\right) \\
&\leq \trans{\left(\sum_{y^{\prime}\in\partial y}v_{y^{\prime} \to y}\right)} (\I - \B)^{-1} \left(\sum_{x^{\prime}\in\partial x}u_{x \to x^{\prime}}\right)
\end{align}
where we have used the assumption $\rho(\B) <1$, so that the series of powers of $\B$ is summable. To write this last equation in a more compact way, recall the definition of the \emph{susceptibility matrix} $\chi\in\mathbb{R}^{n\times n}$, with elements $\chi_{xy} = \E(s_{x}s_{y} )$. Then it holds element-wise that 
\begin{align}
\label{upper_bound_SP}
\chi \leq \P(\I - \B)^{-1}\trans{\Q} + \I_{n\times n}
\end{align}
where $\P,\Q\in\mathbb{R}^{n\times 2m}$ are defined in equation (\ref{P&Q}). 
Note that the addition of an identity matrix ensures that the inequality also holds on the diagonal of $\chi$. This completes the proof of Theorem \ref{th:bound_susceptibility_with_B}. 

\subsection{Proof of Corollary \ref{cor:bound_susceptibility_with_H}}
\label{sec:bound_susceptibility_H}
The corollary will follow from the following lemma, which is of independent interest because it provides another connection between the non-backtracking matrix and the Bethe Hessian, besides the Ihara-Bass formula (\ref{Ihara-Bass}).

\begin{lemma}
\label{lem:B=H}
Let $\B$ be the non-backtracking matrix matrix defined in (\ref{def:B}), and $\H$ the Bethe Hessian defined in (\ref{def:H}). Let $\P,\Q$ be the matrices defined in (\ref{P&Q}). Assume that $\I - \B$ is invertible. Then $\H$ is also invertible, and 
\begin{equation}
\label{B=H}
\P(\I - \B)^{-1}\trans{\Q} +\I_{n\times n} = \H^{-1} \, .
\end{equation}
\end{lemma}

\noindent\emph{Proof:}
The fact that $\H$ is invertible if $\I - \B$ is invertible follows from the Ihara-Bass formula (\ref{Ihara-Bass}).
Take $x\in\mathbb{R}^{n}$, and call $y = (\I - \B)^{-1}\trans{\Q}x$. We wish to show that $\H\P y + \H x =x$. Denoting by $(Py)_{i}$ for $i = 1 \cdots n$ the $i$-th component of the vector $Py$, we have 
\begin{align}
(Py)_{i} &= \sum_{(k \to l)} b_{kl}\delta_{il} y_{k\to l} \\
&= \sum_{k\in \partial i} b_{ki}y_{k\to i} \, .
\label{Py}
\end{align}
$y$ verifies the following equation 
\begin{align}
(\I - \B)y = \trans{\Q}x
\end{align}
so that for all directed edges $(i \to j)$
\begin{align}
y_{i\to j} - \sum_{k\in \partial i\backslash j} b_{ik}y_{k\to i} = x_i
\end{align}
which can be rewritten as 
\begin{align}
&y_{i\to j} - (\P y)_{i} + b_{ij}y_{j\to i} = x_i\, , \\
&y_{j\to i} - (\P y)_{j} + b_{ij}y_{i\to j} = x_j\, ,
\end{align}
where the second equation is for the link $(j \to i)$. Together, these two questions form a closed set of equations allowing to compute $y_{i\to j}$ and $y_{j\to i}$. More precisely, we get 
\begin{align}
y_{i\to j} = \frac{1}{1 - b_{ij}^{2}}\Big( x_{i} + (\P y )_{i} - b_{ij}\left( x_{j} + (\P y )_{j} \right) \Big) \, .
\end{align}
We can now insert this expression in eq. \ref{Py} and get 
\begin{align}
(\P y )_{i} =\sum_{k\in \partial i} b_{ik}y_{k\to i} = \sum_{k\in \partial i} \frac{b_{ik}}{1 - b_{ik}^{2}}\Big( x_{k} + (\P y )_{k} - b_{ik}\left( x_{i} + (\P y )_{i} \right) \Big)
\end{align}
which is equivalent to 
\begin{align}
\left(1 + \sum_{k\in \partial i} \frac{b_{ik}^{2}}{1 - b_{ik}^{2}}\right)(\P y )_{i} - \sum_{k\in\partial i} \frac{b_{ik}}{1-b_{ik}^{2}}(\P y)_{k} = \sum_{k\in\partial i} \frac{b_{ik}}{1-b_{ik}^{2}}x_{k} - \left(\sum_{k\in \partial i} \frac{b_{ik}^{2}}{1 - b_{ik}^{2}}\right)x_{i}
\end{align}
which in matrix form reads 
\begin{align}
\H\P y = - \H x + x
\end{align}
which completes the proof.

\section*{Acknowledgements}
  We thank C.~Borgs, J.~Chayes and A.~Montanari for useful
  discussions. Part of this research has received funding from the European Research Council under the European Union's 7th Framework Program (FP/2007-2013/ERC Grant Agreement 307087-SPARCS).


\bibliographystyle{unsrt}
\bibliography{mybib}

\end{document}